\documentclass[letterpaper,english,reprint, aps]{revtex4-1}
\usepackage[T1]{fontenc}
\usepackage[latin9]{inputenc}
\usepackage[normalem]{ulem}
\usepackage{color}
\usepackage{verbatim}
\usepackage{textcomp}
\usepackage{amsmath}
\usepackage{amssymb}
\usepackage{graphicx}
\usepackage{hyperref}

\makeatletter

\pdfpageheight\paperheight
\pdfpagewidth\paperwidth

\newcommand{\lyxmathsym}[1]{\ifmmode\begingroup\def\b@ld{bold}
  \text{\ifx\math@version\b@ld\bfseries\fi#1}\endgroup\else#1\fi}

\newcommand{\COMMENTED}[1]{}
\newcommand{\CHANGED}[1]{#1}
\newcommand{\DELETED}[1]{}

\makeatother

\usepackage{babel}
\begin{document}
\preprint{APS/123-QED}
\title{Computation of forces and stresses in solids: Towards accurate structural optimization with auxiliary-field quantum Monte Carlo }
\author{Siyuan Chen}
\email{schen24@wm.edu}

\affiliation{Department of Physics, College of William \& Mary, Williamsburg, VA}
\author{Shiwei Zhang}
\email{szhang@flatironinstitute.org}
\affiliation{Center for Computational Quantum Physics, Flatiron Institute, New
York, NY}
\affiliation{Department of Physics, College of William \& Mary, Williamsburg, VA}
\begin{abstract}
The accurate computation of forces and other energy derivatives has  been a long-standing challenge for quantum Monte Carlo methods.
A number of technical obstacles contribute to this challenge.
We discuss how 
these obstacles can be removed
with the auxiliary-field quantum
Monte Carlo (AFQMC) approach. 
AFQMC is a general, high-accuracy, many-body total-energy method for molecules and solids. 
The implementation of back-propagation for pure estimators 
allows direct calculation of gradients of the energy via the Hellmann-Feynman theorem. A planewave basis with norm-conserving pseudopotentials
is used for the study of periodic bulk materials.  
Completeness of the planewave basis minimizes the effect of so-called Pulay terms. The ionic pseudopotentials, which 
can be %
incorporated in AFQMC 
in exactly the same manner as in standard independent-electron methods, 
regulate the force and stress estimators and eliminate any potential divergence of the Monte Carlo variances. 
 The resulting approach allows applications of full geometry
optimizations in bulk materials. %
It also paves the way for 
many-body computations of the phonon spectrum in solids. 
\end{abstract}

\maketitle

\section{\label{sec:intro}Introduction}

Interatomic forces and stresses are two important structural properties
of a solid-state system. As gradients of the potential energy surface
under distortion and deformation, %
they determine %
 the atomic structure and 
 are crucial for geometry optimizations, molecular
dynamics %
simulations, as well as computations of phonon spectrum
and thermodynamic properties, each of which %
constitutes a large and rich set of applications in physics and materials science.

Density functional theory (DFT)~\cite{Hohenberg_PR_1964,Jones_RMP_2015,Becke_JCP_2014,Burke_JCP_2012} has shown incredible success
in computing a wide range of physical properties, including interatomic
forces and stresses. However, in many materials with stronger 
electron-correlation effects, computations based on approximate DFT functionals %
are sometimes not sufficiently accurate to determine structural properties %
~\cite{Gaiduk_JPCL_2015,Cohen_PRB_1990}. 
Many methods are being actively pursued which can better describe electron correlations while 
allowing systematic and realistic calculations to describe molecules and bulk materials. 

Quantum
Monte Carlo (QMC) methods~\cite{Foulkes_RMP_2001} are one class of such methods, which often show a good balance of accuracy and
computational scaling. Indeed QMC methods have seen broad applications in molecules, liquids,
and solids,
and are one of the primary modern tools for post-DFT calculations in electronic structure. However, 
while total energies are straightforward to compute and have been the focal point of QMC methods, 
computations of observables and correlation functions have been less common with QMC in electronic structure.
There have been growing recent efforts to compute properties other than the total energy. Of crucial importance among these are forces and stresses, without which the many-body computations must often rely on
DFT (or experiment, if available) predictions of geometry, and thus cannot be truly predictive in many strongly correlated systems.
\CHANGED{Of course one could compute derivatives by finite difference of the total energy,
including the use of correlated sampling~\cite{Shee_JCTC_2017} and space warp techniques~\cite{Filippi_PRB_2000,Umrigar_IJQC_1989}
for acceleration. However, these have not achieved the desired low computational scaling to allow efficient structural optimization
involving many parameters.}
QMC methods are faced with varying degrees of technical hurdles for \CHANGED{direct, }systematic computations of forces;
to our knowledge no computation of stress tensors has been performed to date.

There are two main forms of QMC methods in electronic structure which have algebraic scaling with system size.
The first includes diffusion Monte Carlo (DMC)~\cite{Reynolds_JCP_1982} and the closely related variational Monte Carlo (VMC)~\cite{McMillan_PR_1965,Ceperley_PRB_1977}, 
which treat the first-quantization Hamiltonian working
in electron coordinate space. 
In  VMC, %
the many-body wavefunction is often explicitly available, %
so  forces
and stresses can in principle be computed directly 
with a modified Hellmann-Feynman estimator \cite{Assaraf_JCP_2003,Assaraf_JCP_2000}.
This has been applied to structural optimizations
\cite{Barborini_JCTC_2012,Guareschi_JCTC_2013} 
and estimations of vibrational properties \CHANGED{in small molecules \cite{Zen_JCTC_2012} and simple solids \cite{Nakano_PRB_2021,Ly_JCP_2022}.}
The accuracy of the computed forces %
are determined by the quality of the variational wave function. 
To date the accuracy has not  consistently reached such a level as to make VMC by itself  a routine post-DFT tool for structural 
optimization, 
especially in strongly correlated systems, although this could change with recent developments of more expressive forms of variational ansatz and better optimization techniques, including with neural networks \cite{Carleo_Science_2017,Pfau_PRR_2020,Hermann_NChem_2020}.
In DMC, the technical hurdles for direct computation of forces and other energy derivatives are more substantial.
In principle evaluation of pure estimators by forward walking is required, which has rarely been performed except for light elements  \cite{Chiesa_PRL_2005,Badinski_PRE_2007}. 
Systematic bias in the mixed estimators, as well as statistical divergences,  must be dealt with
before a general algorithm truly becomes available for structural optimization.
(For a more complete discussion of current state of DMC computations of forces, see for example, Ref.~\cite{Tiihonen_JCP_2021} and references therein.)

The other form of algebraic-scaling QMC methods in electron structure is phase-free
auxiliary-field quantum Monte Carlo (AFQMC)~\cite{Zhang_PRL_2003,Zhang_PRB_1997}, which is the focus of the present work.
AFQMC works in second-quantization, using random walks of non-orthogonal 
Slater determinants in orbital space. 
This formalism provides a non-perturbative, post-DFT method which shares the same Hamiltonian and uses 
much of the same machinery  \cite{Zhang_Handbook2018} as in standard electronic structure. 
The method has had a shorter history of development, but has seen 
growing applications in lattice models of interacting fermions
\cite{Xu_PRR_2022,Qin_PRX_2020},
quantum chemistry
\cite{Motta_WIRES_2018,Shee_JCP_2023},
and solid-state physics~\cite{SC_Density_paper,Purwanto_PRB_2009,Ma_PRL_2015}.
In a number of recent benchmark studies, AFQMC has demonstrated consistently high accuracy for 
total energies in both extended systems~\cite{Motta_PRX_2017,Motta_PRX_2020} 
and  molecules~\cite{Shee_JCTC_2019}, including large transition metal systems~\cite{Williams_PRX_2020}.
 In addition to total energies, expectation values of other observables that do
not commute with Hamiltonian can be computed by a back-propagation (BP) technique \cite{Zhang_PRB_1997,Purwanto_PRE_2004,Motta_JCTC_2017}.
For molecular systems, 
computations of forces using a Gaussian basis set have been performed, with 
geometry optimization %
on small molecules \cite{Motta_JCP_2018}.
In this work we present the computation of  forces and stress tensors in AFQMC using planewaves and pseudopotentials,  to allow full structural optimization of periodic bulk systems.

The remainder of this paper is organized as follows. In Section \ref{sec:theory},
we first briefly review the AFQMC method and the back-propagation technique
for the so-called pure estimators to compute observables.
We then describe the formulation of the atomic forces and stress tensors
within the planewave AFQMC (PW-AFQMC) framework. Section \ref{sec:benchmark} presents systematic benchmarks of
the calculated forces and stresses 
against explicit computations by finite differences, 
which %
validates %
our method and further illustrates its characteristics. In Section \ref{sec:applications},
we show applications %
in two different examples of full structural optimization in solids, one
a geometry optimization of atomic positions in a fixed supercell using
atomic forces, and the other a structural optimization of the cell shape and
size using the stress tensors. 
We then conclude in Section \ref{sec:conclusion}.

\section{\label{sec:theory} Forces and Stresses in Plane-wave AFQMC}

\subsection{\label{ssec:review-AFQMC}Basic formalism of AFQMC}

AFQMC~\cite{Zhang_PRL_2003,Zhang_PRB_1997} approaches the many-body ground state of a system
with imaginary time propagation $\lim_{N\to\infty}e^{-N\Delta\tau H}|\Psi_{\mathrm{T}}\rangle\to|\Psi_{0}$$\rangle$,
where 
$H$ is the many-body Hamiltonian whose
ground state $|\Psi_{0}\rangle$ is targeted, %
$|\Psi_{\mathrm{T}}\rangle$ is a %
trial wave function
that is not orthogonal with $|\Psi_{0}\rangle$.
The propagation is separated into $N$ steps, each of which of imaginary time
length $\Delta\tau$, making the propagation an iterative process.
The size of the time-step $\Delta\tau$ must be chosen to be sufficiently small 
to minimize commutator errors, known as Trotter errors.
The algorithm takes the form of an open-ended random walk, such that there is little restriction on $N$, which typically 
takes very large values.

An interacting electronic 
Hamiltonian, such as the ones in electronic structure under the 
Born-Oppenheimer approximation,
contains one-body and two-body terms. %
Propagating with the exponential of
one-body terms takes a %
Slater determinant to another Slater determinant~\cite{ThoulessTheorem}. %
Two-body propagators, which do not preserve the form of a Slater determinant,
are treated %
in AFQMC via the %
Hubbard-Stratonovich transformation~\cite{Stratonovich_DAN_1957,Hubbard_PRL_1959}:
\begin{equation}
e^{-\frac{\Delta\tau}{2}\lambda\hat{v}^{2}}=\frac{1}{\sqrt{2\pi}}\int_{-\infty}^{\infty}dxe^{-\frac{1}{2}x^{2}}e^{x\sqrt{-\Delta\tau\lambda}\hat{v}}\,.
\end{equation}
This formula rewrites the propagator of any two-body Hamiltonian term, after it has been expressed 
in the form of the sum of squares of one-body operators: $H_{2}=\sum_{i}a_{i}\hat{v}_{i}^{2}$, 
into an integral of one-body propagators.
The
integral over the  auxiliary fields,  $\{x_{i}\}$, can be then evaluated with Monte Carlo.

The iterative process of imaginary time propagation then transforms
into a random walk process of a population of Slater determinants (walkers) $\{|\Phi_k^{(n)}\rangle\}$,
where $n$ indicates the imaginary-time step count, and $k$ is an index of the random walker at each time $n$. 
Each walker $|\Phi^{(n)}\rangle$ is a Slater determinant propagated from 
the initial determinant, and is dependent on  its specific path history in auxiliary-field (AF) space, $\{ \{x_i\}^{(n)}, \{x_i\}^{(n-1)}, \cdots, \{x_i\}^{(1)}\}$
(omitting the walker index $k$).  
The wavefunction at each step is represented
by a weighted average of all the random walkers %
at that step, 
$|\Psi^{(n)}\rangle\propto \sum_{k} |\Phi_k^{(n)}\rangle/\langle \Psi_\mathrm{T}|\Phi_k^{(n)}\rangle$
and it approaches
the ground state after a sufficiently large number of %
steps $n>n_{\rm eq}$. The value $n_{\rm eq}$ depends on $|\Psi_\mathrm{T}\rangle$ and the system, and is such that $n_{\rm eq}\Delta\tau $ allows the imaginary-time projection from $|\Psi_\mathrm{T}\rangle$ to reach $|\Psi_0\rangle$ within the desired statistical accuracy. 
After convergence, both ensemble and time averages together 
give a representation of the ground-state wave function, 
$|\Psi_0\rangle\propto \sum_{n>n_{\rm eq}} |\Psi^{(n)}\rangle$, 
whose statistical accuracy can be improved with increasing sample size, following 
the behavior dictated by the central limit theorem. 
The actual AFQMC algorithm is augmented by several additional ingredients, 
including importance sampling (which is embedded in the form of $|\Psi^{(n)}\rangle$ we used above), 
and the use of a force bias in proposing Monte Carlo moves to improve efficiency~\cite{Purwanto_PRE_2004}, as well 
as the phaseless approximation to control the phase problem~\cite{Zhang_PRL_2003}.

The open-ended random walk scheme yields a form %
to conveniently evaluate observables
that commute with the Hamiltonian, %
using the \emph{mixed estimator}. For example, the total energy
can be computed through 
\begin{equation}
\langle H\rangle = E=\frac{\langle\Psi_{\mathrm{T}}|H|\Psi_{0}\rangle}{\langle\Psi_{\mathrm{T}}|\Psi_{0}\rangle}\,,
\label{eq:E_mix}
\end{equation}
for which we only need to propagate one side in the estimator, the ket. %
The numerator and the denominator can be computed with the random walk averages, and 
the final estimator for the energy involves weighted averages of ``local energies" 
of the form $E_L(\Phi_k^{(n)})=\langle \Psi_\mathrm{T}|H|\Phi_k^{(n)}\rangle/
\langle \Psi_\mathrm{T}|\Phi_k^{(n)}\rangle$.

For observables which do not commute with the Hamiltonian,
computations with the 
mixed estimator in Eq.~(\ref{eq:E_mix}) will incur a bias.
A more accurate calculation will require
propagation of the bra  %
 $\langle\Psi_{\mathrm{T}}|$ to the ground
state as well, the so-called \emph{pure estimator}. 
This is nominally not difficult to achieve. For example one could sample an entire path of AF for a fixed length of imaginary time with the generalized 
Metropolis algorithm \cite{Zhang_LectureNotes_2019}. However, this approach 
would cause ergodicity problems when a constraint needs to be imposed 
along the path to control the sign or phase problem. 
In the open-ended 
random walk formulation with importance sampling and constraint, as mentioned above, the projection of the left-side requires the back propagation (BP) scheme \cite{Zhang_PRB_1997,Purwanto_PRE_2004,Motta_JCTC_2017} referred to earlier.

We observe that
\begin{equation}
\langle O\rangle\simeq\frac{\langle\Psi_{\mathrm{T}}|e^{-m\Delta\tau H}\hat{O}e^{-n\Delta\tau H}|\Psi_{\mathrm{T}}\rangle}{\langle\Psi_{\mathrm{T}}|e^{-(m+n)\Delta\tau H}|\Psi_{\mathrm{T}}\rangle}\,,
\label{eq:pure-est}
\end{equation}
where $\langle O\rangle$ approaches %
the ground-state expectation as $m,n\to \infty$.
The denominator can be viewed as an overlap of the trial wave function
with a propagation of $(m+n)$ steps. If we choose to remember the
last $m$ steps of the AFs %
and propagate 
$\langle\Psi_{\mathrm{T}}|$
back with 
the corresponding
one-body operators in reverse order, we obtain an %
estimate of the propagated bra
$\langle\Psi_{0}|\simeq\langle\Psi_{\mathrm{T}}|e^{-m\Delta\tau H}$.
This is %
the basic idea of BP in %
AFQMC, %
which allows a 
seamless integration of the backward projection with the importance sampling scheme applied in the forward direction.
The BP scheme has been applied widely in calculations on lattice models
of strong correlations \cite{Vitali_PRB_2016,Qin_PRX_2020,Xu_PRR_2022}.  
An additional bias arises in BP %
because of the reversal of the direction in which the constraint is applied. 
Such biases are generally much
smaller than the mixed-estimator bias for observables that do not commute with the Hamiltonian, 
but can be larger than that of the purely variational estimator (which is often hard to compute) \cite{Purwanto_PRE_2004}. 
We apply  the recently proposed path-restoration technique~\cite{Motta_JCTC_2017},
which can further mitigate the BP %
bias. Our implementation 
 of the BP %
 scheme in planewave AFQMC is discussed in more detail in Ref.~\cite{SC_Density_paper}. 
 For the purpose of the present work, the most important aspect to note is
 that Eq.~(\ref{eq:pure-est}) is reduced to weighted averages of local estimators of the form
 \begin{equation}
\langle O\rangle^\mathrm{BP}_k
\equiv\frac{\langle\bar\Phi_k^{(m)}|\hat{O}|\Phi_k^{(n)}\rangle}
{\langle\bar\Phi_k^{(m)}|\Phi_k^{(n)}\rangle}\,,
\label{eq:bp-local-est}
\end{equation}
where $k$ labels a walker which survives through the $(m+n)^{\rm th}$ step of the random walk, $|\Phi_k^{(n)}\rangle$ is the parent walker of $k$ back 
in the $n^{\rm th}$ step, and $\langle\bar\Phi_k^{(m)}|$ is the back-propagated bra Slater determinant.
The weighted average over $k$ yields the Monte Carlo estimate of the expectation value of $O$ given in  Eq.~(\ref{eq:pure-est}).

Any one-body  operator $O = \sum_{uv} A_{uv} c^\dagger_u c_v $
or two-body operator $O = \sum_{pqrs} V_{pqrs} c^\dagger_p c^\dagger_q c_s c_r$, or their linear combinations,
can be computed with the above approach. 
The estimators $\langle c^\dagger_u c_v \rangle$ and 
$\langle c^\dagger_p c^\dagger_q c_s c_r \rangle$
are the one-body and two-body reduced density matrices (1rdm, 2rdm) 
$\mathcal{G}_{uv}$ and $\mathcal{G}_{pqrs}$, respectively.
Computation of $\langle O\rangle$ %
can therefore be thought of as computing the 1rdm's and 
2rdm's (which can be obtained via Wick's theorem~\cite{WickTheorem,Motta_WIRES_2018}),
and then multiply them with the corresponding coefficients $A_{uv}$ and $V_{pqrs}$.
This straightforward approach is ineffective with the plane wave basis, where the number of basis 
functions is much larger than with a localized basis set choice. 
As such, naive implementations would lead to
\CHANGED{
large storage ($\mathcal{O}(N_\mathrm{PW}^2)$ for the 1rdm)
and computational costs ($\mathcal{O}(N_\mathrm{PW}^3)$ for operations like
$\mathrm{Tr}(A_{uv} \mathcal{G}_{uv})$).
}

Instead we take a different approach in planewave AFQMC. Recall
\begin{equation}
\mathcal{G}_{uv} = \mathrm{Tr} [({\Phi^\dagger \Psi})^{-1}{\Phi^\dagger {\mathcal E}_{uv} \Psi}] 
= [\Psi ({\Phi^\dagger \Psi})^{-1} \Phi^\dagger]_{vu}\,,
\end{equation}
\noindent
where $\Psi$ and $\Phi$ are the matrix form of the ket and bra Slater determinants, and ${\mathcal E}$ is a matrix with only one nonzero element ${\mathcal E}_{uv}=1$.
We store the intermediate matrix
$\Theta = \Psi ({\Phi^\dagger \Psi})^{-1}$,
which only requires a memory of $\mathcal{O}(N_\mathrm{PW}N_e)$. 
The 1rdm is conveniently restored from $\Theta$ and the bra determinant:
\begin{equation}
\mathcal{G}_{uv} = \sum_{t=1}^{N_e} \Theta_{vt} (\Phi^\dagger)_{tu}\,.
\label{eq:Theta-Phi}
\end{equation}
The use of fast Fourier transforms (FFTs) and convolutions lead to efficient evaluations.
For example, the local part of the electron-ion interaction (see next section for further details):
\begin{equation}
V_{\mathrm{ei}}^\mathrm{L}=\sum_{\mathbf{Q}\neq \mathbf{0}}v_{\mathrm{ei}}^\mathrm{L}(\mathbf{Q})\rho(\mathbf{Q})\,,
\end{equation}
\noindent
with $\rho(\mathbf{Q}) \equiv \sum_\mathbf{G} c^\dagger_\mathbf{G} c_{\mathbf{G}+\mathbf{Q}}$ the %
``density operator'' in $\mathbf{Q}$-space, is given as
\begin{equation}
\mathrm{Tr}[V_{\mathrm{ei}}^\mathrm{L} \mathcal{G}] 
= \sum_{t=1}^{N_e}\sum_{\mathbf{G}} \Phi^\dagger_{t\mathbf{G}} 
\sum_\mathbf{Q} v_{\mathrm{ei}}^\mathrm{L}(\mathbf{Q}) \Theta_{\mathbf{G}+\mathbf{Q},t} \,,
\end{equation}
which involves a convolution in the form of 
$(A\star B)_\mathbf{q} = \sum_\mathbf{p} A_\mathbf{p} B_{\pm \mathbf{p}+\mathbf{q}}$, 
that is conveniently computed with FFTs and inverse FFTs on the plane-wave grid, and only has a complexity of $\mathcal{O}(N_e N_\mathrm{PW} \log N_\mathrm{PW})$. The sum on the outer layer also only requires a complexity of $\mathcal{O}(N_e^2 N_\mathrm{PW})$.

\subsection{\label{ssec:forces-stress} The computation of forces and stresses in planewave AFQMC}

With BP and path restoration, pure expectation 
values of observables can be computed. %
This allows us to then apply the Hellmann-Feynman (HF) theorem to 
compute the expectation values of the derivatives of the Hamiltonian directly.
Computation of AFQMC forces and stresses are then available,
which are given via the HF theorem as 
expectations of the derivatives of the Hamiltonian. 

In the plane-wave basis,
the second-quantized Born-Oppenheimer Hamiltonian $H$
can be %
written as a sum of following components~\cite{Suewattana_PRB_2007}:
\begin{equation}
H=K+V_{\mathrm{ei}}+\gamma_{\mathrm{Ewald}}+V_{\mathrm{ee}}\,,
\label{eq:plane-wave-H}
\end{equation}
\noindent which are the kinetic energy, 
the electron-ion interaction 
(represented by pseudopotentials),
the Ewald energy (a system-related constant coming from the interaction of the ions, including
with their images due to the periodic cell), and the electron-electron
interaction, respectively.
A kinetic energy cutoff $|\mathbf{G}|^{2}<E_{\mathrm{cut}}$
is imposed on the plane waves, limiting the total number of plane
waves to a finite number $N_{\mathrm{PW}}$.
The pseudopotential can be separated into local (L) and nonlocal (NL) components~\cite{Suewattana_PRB_2007}:  %
\begin{equation}
V_{\mathrm{ei}}=\sum_{\mathbf{Q}\neq \mathbf{0}}v_{\mathrm{ei}}^\mathrm{L}(\mathbf{Q},\{\vec{\tau}\})\rho(\mathbf{Q})+
\sum_{\mathbf{G},\mathbf{G'}}v_{\mathrm{ei}}^\mathrm{NL}(\mathbf{G},\mathbf{G}',\{\vec{\tau}\})c_{\mathbf{G}}^{\dagger}c_{\mathbf{G}'}\,,
\label{eq:plane-wave-PSP}
\end{equation}
where  $\mathbf{G}$  and  $\mathbf{G'}$ are planewaves within the cutoff $E_{\rm cut}$, $\mathbf{Q}\equiv \mathbf{G'}-\mathbf{G}$, 
the operator $\rho$ is the Fourier transform of the real-space electronic density, 
and ${\vec \tau}$ denotes the positions of ions. We have omitted the spin index in the operators.
The electron-electron interaction is $V_\mathrm{ee} =V^\mathrm{C} + N \xi$, where 
the constant second term (with $N$ being the number of electrons) is similar to the Ewald term from the ions and can be treated together with the latter for convenience, and
\begin{equation}
V^\mathrm{C} \equiv \frac{4\pi}{\Omega}\sum_{pqrs}\sum_{\mathbf{Q}\neq \mathbf{0}}\frac{1}{|\mathbf{Q}|^2} c^\dagger_p c^\dagger_q c_s c_r\,,
\label{eq:Vee}
\end{equation}
\noindent
where each of the indices $p$, $q$, $r$, $s$ denotes a combination of  plane-wave vector $\mathbf{G}$ and spin $\sigma$.
In Eq.~(\ref{eq:Vee}), momentum conservation $\mathbf{G}_r+\mathbf{G}_s=\mathbf{G}_p+\mathbf{G}_q$ and spin invariance  $\sigma_r=\sigma_p$, $\sigma_q=\sigma_s$ 
are imposed,  and a sum over the spin indices is implicit.

Interatomic forces are derivatives of the total energy with respect to ion positions $\{\vec{\tau}\}$,
which are only present in the pseudopotential and ion-ion Ewald energy. 
From Hellmann-Feynman theorem:
\begin{equation}
F_{ia} = -\frac{\partial E}{\partial \tau_{ia}} = \langle \Psi_0 | -\frac{\partial H}{\partial \tau_{ia}} | \Psi_0 \rangle \equiv 
\langle \Psi_0 | \hat{F}_{ia} | \Psi_0 \rangle \,,
\label{eq:HF-theorem}
\end{equation}
where $i$ marks each atom and $a$ marks each of the 3 Cartesian directions.
The force observable that will replace $\hat{O}$ in Eq.~\ref{eq:bp-local-est} is therefore written as:
\begin{equation}
\hat{\mathbf{F}} = \mathbf{F}_\mathrm{Ewald} + \hat{\mathbf{F}}_\mathrm{ei}\,,
\label{eq:force-obs}
\end{equation}
where
the Ewald force $\mathbf{F}_\mathrm{Ewald}$ is a constant %
\cite{Martin_ES_2020}. 
For the electron-ion contribution, 
the dependence on %
ion positions is %
only %
in the coefficients
$v_{\mathrm{ei}}$, as seen in Eq.~(\ref{eq:plane-wave-PSP}).
The computation of the electron-ion forces therefore requires only a replacement
of the coefficients  $v_{\mathrm{ei}}$ in the total energy computations by $-\partial v_{\mathrm{ei}}/\partial \tau_{ia}$.
As all dependencies of $\{\vec{\tau}\}$ in $v_\mathrm{ei}$ are in the form of structure factors (of the form $e^{i \mathbf{G} \cdot \vec{\tau}}$ -- see Appendix~\ref{app:psp-force}), computations of $-\partial v_{\mathrm{ei}}/\partial \tau_{ia}$ are straightforward.
There is no dependence of the ion positions in the plane-wave basis, hence no Pulay terms from the basis set here.

The stress tensors $\sigma_{ab}$ are derivatives of total energy with respect to a %
strain %
$\epsilon_{ab}$,
which describes the deformation $\mathbf{U}$ of any crystal point with respect to its (Cartesian) coordinates $\mathbf{X}$,
$\epsilon_{ab} = \partial U_a / \partial X_b$.
The stress tensor is then defined as
\begin{equation}
\sigma_{ab}=-\frac{1}{\Omega}\frac{\partial E}{\partial\epsilon_{ab}} \,,
\label{eq:stress-def}
\end{equation}
where $\Omega$ is the supercell volume.  
As the strain tensor is transpose symmetric, so is the stress tensor. 
Because of statistical errors,  this symmetry only holds in a statistical sense in AFQMC.
We apply an explicit symmetrization of the stress tensor after the AFQMC calculation:  
$\bar{\sigma}_{ab}\equiv ({\sigma_{ab}+\sigma_{ba}})/2$.

Unlike forces, the Hamiltonian terms are not directly 
dependent on the strain
tensor so a chain rule has to be applied through all real-space and
reciprocal-space vectors, as well as the lattice volume. This is based
on a list of transforms under strain: $\mathbf{r}_{a}\to\sum_{b}(\delta_{ab}+\epsilon_{ab})\mathbf{r}_{b}$,
$\mathbf{k}_{a}\to\sum_{b}(\delta_{ab}-\epsilon_{ab})\mathbf{k}_{b}$, and
$\Omega\to(1+\Sigma_{a}\epsilon_{aa})\Omega$, where $\mathbf{r}$ and $\mathbf{k}$ represent, respectively,
\textit{any} real- and reciprocal-space vectors in the Hamiltonian. The observable to 
 evaluate by Eq.~\ref{eq:bp-local-est} is therefore:
\begin{equation}
\hat{\sigma}_{ab} = -\frac{1}{\Omega}(\sum_{\mathbf{r},c} \delta_{ac} r_b \frac{\partial \hat{H}}{\partial r_c} - \sum_{\mathbf{k},c} \delta_{ac} k_b \frac{\partial \hat{H}}{\partial k_c} + \delta_{ab}\Omega \frac{\partial \hat{H}}{\partial \Omega}) \,.
\label{eq:stress-HF}
\end{equation}
Every term in the Hamiltonian in Eq.~(\ref{eq:plane-wave-H}) is affected by the change of the space metric, which
means a derivative is needed for each. %
We write it as
\begin{equation}
\hat{\sigma} = \hat{\sigma}_\mathrm{K} +  \hat{\sigma}_\mathrm{ei} + \hat{\sigma}_\mathrm{Ewald} + \hat{\sigma}_\mathrm{ee}\,.
\end{equation}
\noindent
The kinetic and Ewald terms are formally the same as in the corresponding DFT calculations 
~\cite{Martin_ES_2020}. 
Dependencies on $\mathbf{G}$ and $\mathbf{Q}$ arise in 
 the electron-ion contribution 
in Eq.~(\ref{eq:plane-wave-PSP}),  
which result in derivatives of the pseudopotential function and the spherical harmonics 
(see Appendix~\ref{app:psp-stress} for details.)
For the electron-electron interaction, the contribution to the stress from the Ewald term is readily available
(by setting $Z_i\to-1$\CHANGED{$,\vec{\tau}\to\mathbf{0}$} in~\cite{Martin_ES_2020}).
The remaining contribution, from Eq.~(\ref{eq:Vee}), is %
\begin{equation}
\sigma_{ab}^\mathrm{C} = \frac{\delta_{ab}}{\Omega}V^\mathrm{C} - \frac{8\pi}{\Omega^2}\sum_{pqrs}\sum_{\mathbf{Q}\neq \mathbf{0}}\frac{Q_a Q_b}{|\mathbf{Q}|^4} c^\dagger_p c^\dagger_q c_s c_r \,,
\end{equation}
\noindent
where the second term can be computed similarly to the first term which is already present in the total energy calculation.

\CHANGED{
We comment on the computational cost of forces and stresses, compared with a total-energy computation.
BP is performed occasionally in AFQMC, so it only adds a small additional cost. The computational scaling of
BP is also the same as energy computations; in both cases the major cost is in estimating 1rdms.
The computational scaling for forces and stresses is therefore the same as total-energy-only computations,
with an additional prefactor ($\sim$1.2$\times$ in the examples we tested in this work).
}

\subsection{\label{ssec:errors} Sources of errors and their mitigation}

At the top level, the formalism we have presented for computing atomic forces and stress tensors have two 
sources of systematic errors. The first is from the phaseless constraint of AFQMC, 
which controls the sign or phase problem. 
In other words, the ground-state wave function sampled from the AFQMC, $|\tilde \Psi_0\rangle$, deviates from the 
exact $|\Psi_0\rangle$. 
This bias is reflected in the computed total energy (from the mixed estimator), and is generally very small, 
as seen through many studies and in the large body of benchmark results \cite{Williams_PRX_2020,Motta_PRX_2017}. 
Additional reduction of the systematic errors can be achieved by better trial wave functions or the use of self-consistent constraints \cite{Shi_JCP_2021}.

The other source of error is the BP bias. %
If $|\tilde \Psi_0\rangle$ can be used on both sides to compute a variational estimate of 
$\langle O\rangle$, the result is expected to be of a quality consistent  with the total energy \cite{Purwanto_PRE_2004}.
However, we cannot do this very efficiently in general, and instead use the BP approach,
in which the backward walker paths in Eq.~(\ref{eq:pure-est})
do not satisfy the rigorous constraining sign or gauge condition, which is imposed
in the forward-propagating direction \cite{Carlson_PRB_1999}. 
This bias is mitigated (but not fully suppressed)
by the path-restoration scheme, as discussed and illustrated in \cite{Motta_JCTC_2017}.
The accuracy of the BP result can still be below that expected from the total energy. 
One very useful way to quantify this error is via explicit calculations of $\langle O\rangle$,
by finite difference using multiple total energy calculations. (This approach has seen many 
applications in lattice models \cite{Qin_PRX_2020}.) The benchmark results below in Sec.~\ref{sec:benchmark} are precisely in this mode, and the excellent agreement between our direct results and the target finite-difference values indicates negligible BP error.

Other sources of errors are present but can be systematically removed. These for example include
Trotter errors, population control bias (both of which are also present in total-energy-only calculations), and BP equilibration time bias, all of which can be handled 
 in standard ways \cite{Motta_WIRES_2018}. 
 
 We comment on two other errors which require a bit more attention for forces and especially stress tensors, namely finite-size error and residual basis set error. 
First, AFQMC computations are performed in finite systems, and the results must be extrapolated to the thermodynamic limit for bulk systems. This applies to the forces and stresses we compute as well. 
To help reduce finite-size effects, we apply a post-processing correction from a finite-size DFT functional parameterized in 
Ref.~\cite{Kwee_PRL_2008} (referred to as KZK in the literature). The KZK finite-size correction is for the total energy. 
Since forces and stress tensors are both energy derivatives, we can in principle apply a post-processing to them in the same way as to the total energy~\cite{Kwee_PRL_2008}.
However, for the stress tensors, coefficients appearing in the KZK finite-size functional are dependent
on the lattice volume, whose 
derivatives %
must therefore be accounted
for. A simple way to treat this problem and avoiding additional %
Pulay terms %
is to use the finite-difference KZK stress
$\sigma_{\mathrm{KZK},ab}=-\Delta E_{\mathrm{KZK}}/(\Omega\Delta\epsilon_{ab})$.
After that, the usual way of finite-size correction $\sigma_{\mathrm{QMC}}^{\infty}=\sigma_{\mathrm{QMC}}^{\mathrm{FS}}-\sigma_{\mathrm{KZK}}+\sigma_{\mathrm{DFT}}^{\infty}$
can be applied.

The second point worth noting concerns finite basis set errors, or rather the (lack of) balance between the plane wave basis sets in different supercells.
As mentioned, 
the plane-wave basis set, which is independent of ionic positions in the supercell, has essentially no finite basis error for force
calculations within a fixed supercell. 
It does depend on the space metric, and the number of plane waves varies with the supercell size. A Pulay term thus arises for 
stress tensors.
We find this Pulay term
to be minimal (``kbar''-level) for a suitable PW cutoff.
If a higher accuracy is desired,
common solutions from DFT, such as increasing or smoothing the cutoff~\cite{Bernasconi_JPCS_1995}, %
can be adopted straightforwardly in PW-AFQMC and works well. 
An even simpler scheme, in the spirit of KZK,
is to correct QMC results with the
corresponding DFT cutoff error: $\sigma_{\mathrm{QMC}}^{E_{\mathrm{cut}}=\infty}\approx\sigma_{\mathrm{QMC}}^{E_{\mathrm{cut}}}-\sigma_{\mathrm{DFT}}^{E_{\mathrm{cut}}}+\sigma_{\mathrm{DFT}}^{E_{\mathrm{cut}}=\infty}$. 
Although approximate, this scheme works well for moderately correlated materials.

\section{\label{sec:benchmark}Benchmark and illustration }

To validate our formalism and implementation, and test the accuracy of  
 force and stress computed with PW-AFQMC,
we performed a number of benchmark calculations.
 We compare the forces and stress tensors computed directly by the approach outlined in Sec.~\ref{ssec:forces-stress} with 
 the corresponding finite difference results obtained from AFQMC total energies. The comparison is made in 
 a finite system under identical conditions. The total energy calculations are fully converged with respect to any systematic errors except for the phaseless error, which is expected to be negligibly small in these systems \cite{Malone_PRB_2020}. We ensure that the error from finite difference is smaller than the statistical error in the reference data. Trotter step
sizes are extrapolated to zero from three separate finite step-size computations.
 As discussed in Sec.~\ref{ssec:errors}, this comparison thus quantifies all the errors in the forces and stress tensors except that  from the phaseless constraint.  

\begin{figure}
\includegraphics[width=1\linewidth]{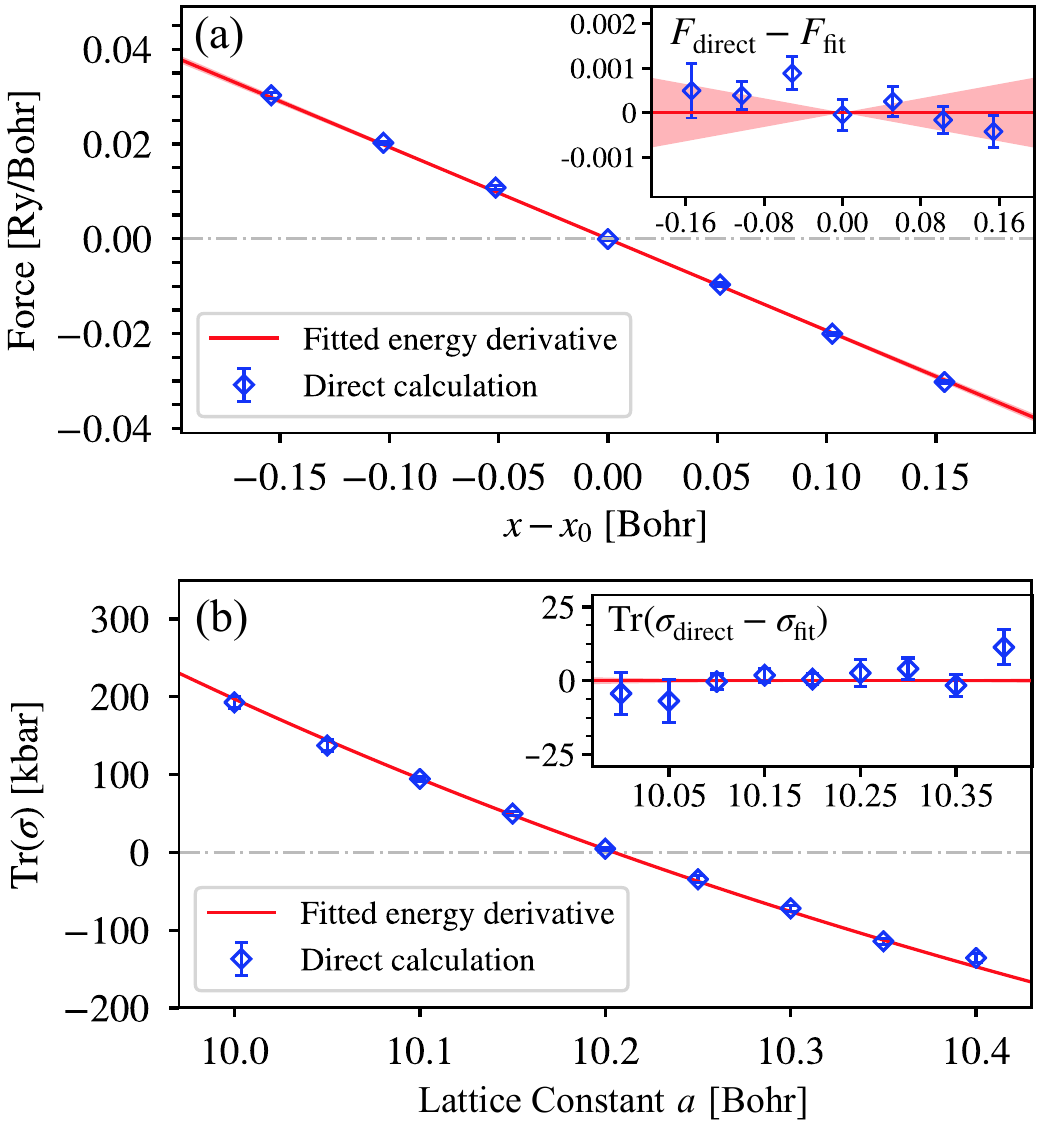}\caption{\label{fig:f-s-benchmark}Benchmark of the computed forces (top panel) and stress tensors (bottom panel) in the Si
diamond structure.
 Forces/stresses directly computed by
AFQMC are shown by blue diamonds with error bars, and the reference data, from 
differentiating the AFQMC total energies, are shown by the red solid curve with
error bar as shades. The insets show a zoomed view of the difference between the
two. In (a), the horizontal axis gives the displacement of one atom along one direction. 
In (b) it is the lattice constant as the cell is varied. 
}
\end{figure}

We consider a diamond-structured Si in the primitive face-centered cubic (FCC) cell.
To benchmark forces, we displace one Si atom along the Cartesian  $x$-axis
of the cell. 
We compare the directly computed forces with the reference result from total energies across a range of 
displacement,  
from -1.5\% to 1.5\% of the experimental lattice constant (10.263
Bohr) with a 0.5\% step interval. 
To obtain the reference data, we compute the total energies with AFQMC across a wider
range (-2\% to 2\% of the lattice
constant). 
We then fit the computed total energy to the quadratic function $E=\frac{1}{2}kx^{2}+E_{0}$.
(We have verified that this form is sufficient, as expected for the vicinity of the equilibrium.) 
The fit is performed in a stochastic way to account for the statistical error bars in the computed total energy: 
a value is selected randomly %
at each data point %
from a Gaussian distribution centered at the mean, with variance given by the Monte Carlo error bar;
the set of values for the entire 
displacement range forms a ``sample'' which can be fitted to
obtain a $\{k,E_{0}\}$; a large number of samples are used to estimate
the value and uncertainty of $\{k,E_{0}\}$ through the sample average
and standard deviation. As a positional derivative of the total energy,
the fitted force is then given by %
 $F=-kx$, with statistical uncertainty from the value of $k$. 
 (This is seen in the linearly growing statistical uncertainty in the reference data in the inset of Fig.~\ref{fig:f-s-benchmark}(a).)
 This reference %
 force $F_{\mathrm{fit}}$ is then compared with
the force directly computed from AFQMC using the algorithm in Sec.~\ref{ssec:forces-stress}, $F_{\mathrm{direct}}$. 
As shown in
Fig.~\ref{fig:f-s-benchmark}(a), excellent agreement is seen across the entire range. 

To benchmark the computed stress tensors, we proceed in a similar fashion, by deforming
the %
lattice to vary the cell volume and shape, and computing
the derivatives of the equation of state to obtain reference data. 
Here we show an example on %
the
diagonal stress terms, which are associated with lattice volume changes.
We use the same silicon structure, varying the lattice constant $a$ %
around
the experimental equilibrium value 
and calculating
the total energy %
for a range of lattice constants  (9.8 to 10.6 Bohr).
Similar to the force benchmark, this range is larger than that targeted in the direct stress calculations, in order to obtain a reliable fit across the range of the benchmark. %
We then fit the computed equation of state with the Murnaghan equation
\cite{Murnaghan_PNAS_1944} following the same stochastic procedure described above, and obtain estimates of the %
the four free parameters $\{E_{0},V_{0},K_{0},K_{0}'\}$ and their statistical uncertainties. %
Noting that
\begin{equation}
-\frac{3a}{\Omega}\frac{\partial E}{\partial a}=-\sum_{i=1}^{3}\frac{1}{\Omega}\frac{\partial E}{\partial\epsilon_{ii}}=\mathrm{Tr}[\sigma]\,,
\end{equation}
we can evaluate the strain derivative in the middle by the left-hand side from the Murnaghan equation with the fitted parameters, and compare  it 
with the trace of the directly computed stress matrix on the right-hand side.
The results are presented in Fig.~\ref{fig:f-s-benchmark}(b).
 In the main graph, Pulay corrections have been applied to both sets of data.
 The position in $a$ where either result intercepts 0 shows a small discrepancy from the experimental equilibrium lattice constant.
This arises from residual finite-size error (which should vanish when
extrapolated to the thermodynamic limit) and has no effect for the purpose here.
Excellent agreement is again seen between the computed stress and the benchmark data. %

\section{\label{sec:applications}Applications in Geometry Optimization}

\begin{figure}
\includegraphics[width=1\linewidth]{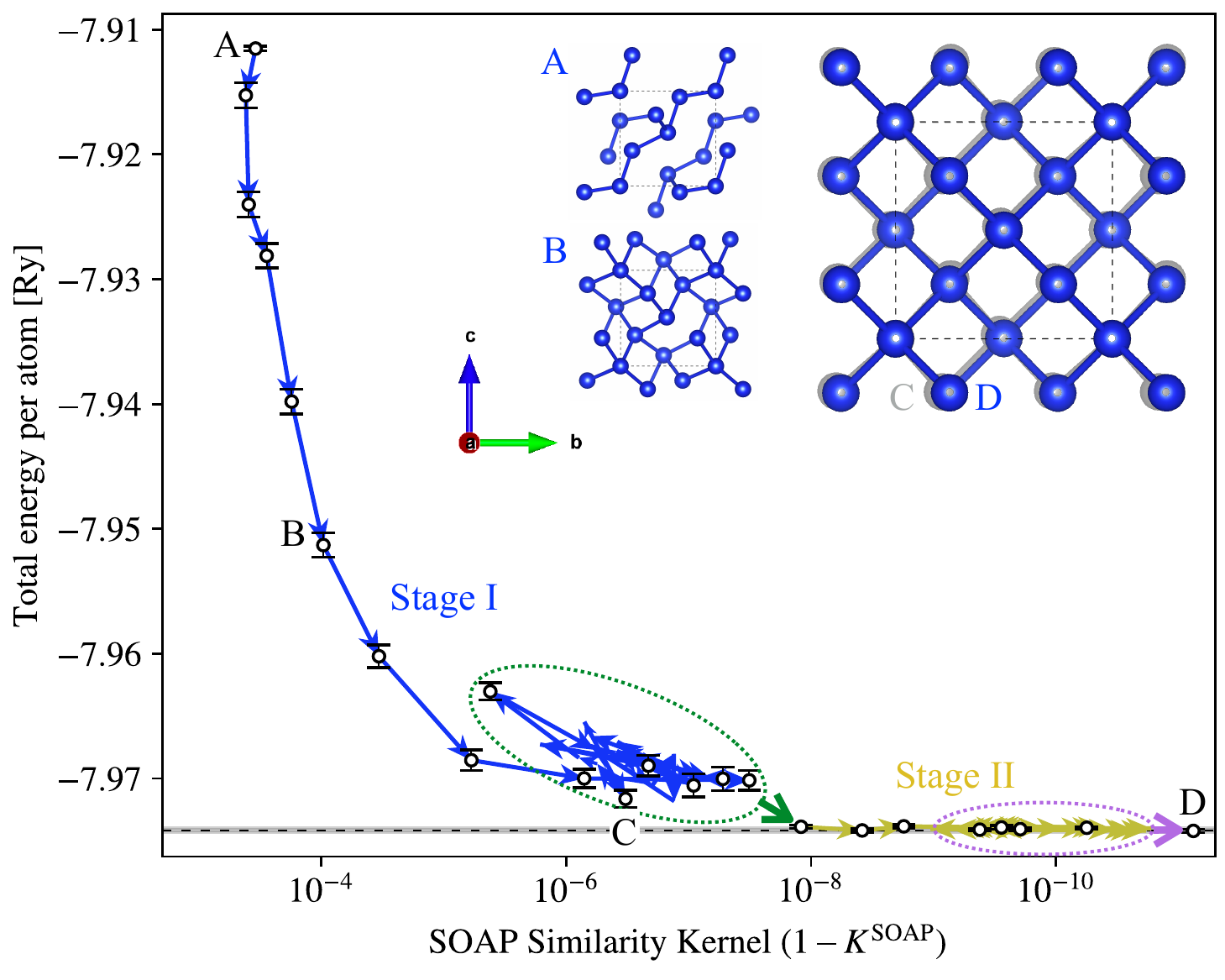}\caption{\label{fig:force-FSSD}
Optimization of all atomic positions in a supercell of diamond %
Si. The starting structure (A) is a 50:50 mix of \CHANGED{atomic positions} in
diamond and $\beta$-tin structures, placed inside a supercell of equilibrium volume of the diamond structure. %
The target is the global minimum diamond structure (D). 
The X axis shows the SOAP similarity kernel~\cite{Bartok_SOAP_2013}. %
The Y axis shows the AFQMC total energy per Si atom.
The scale of
statistical uncertainty in the %
energy is indicated by the error bars 
at selected steps. %
The black dashed line shows the %
 energy computed at the target diamond structure,
with the gray shades indicating the statistical error.
The insets A-D show the atomic positions in the $y$-$z$ plane for four steps along the optimization trajectory, as indicated.
Structures C and D are very close %
and %
are shown as overlapping images. %
}
\end{figure}

The ability to compute accurate force and stress from AFQMC can potentially enable many
applications. One of these %
is geometry optimization. %
A full degree-of-freedom (DOF) geometry optimization
is possible when we have both forces and stresses available. Interatomic
forces allow for optimizations in atom positions, and the stress tensors
allow for optimizations of the lattice volume and shape. 
Here as a first test, we apply these capabilities to   two different bulk systems: Si and aluminium nitride (AlN).

The computed forces and stresses can be fed into any optimization routine for structural optimization. 
Here we use an optimization algorithm that we recently developed~\cite{SC_GeoOpt_paper},  called
FSSD$\times$SET (fixed step-size descent with staged error targeting). In a series of tests,
in which we emulated 
forces and stresses computed from QMC (or any other methods which might contain stochastic noise) by adding synthetic noise to the corresponding DFT results, we studied the efficiency and effectiveness of commonly applied structural optimization algorithms, including some of the latest machine learning optimization methods. We found that the FSSD$\times$SET approach 
consistently performed efficiently and robustly under realistic conditions. In the test examples below, we thus apply this algorithm in combination with forces or stresses computed from AFQMC to realize fully \textit{ab initio} many-body structural optimizations. 

We first perform a geometry optimization of atomic positions %
in bulk Si. 
We consider a cubic supercell with the experimental equilibrium lattice constant of  $a=10.263$\,Bohr.
The initial positions of the atoms are a 50:50
mix of their fractional coordinates
in the diamond structure and the $\beta$-tin structure (under strain). %
Fig.~\ref{fig:force-FSSD} illustrates how the system, under PW-AFQMC optimization, transforms
into the diamond structure.
Arrows connect subsequent steps, and in this optimization run, the SET includes two stages, marked by two different colors.
At the beginning of the optimization (the first stage), the total energy drops quickly 
and in a few steps the atoms change from their initial positions (depicted in A) to form a pattern that looks like a distorted diamond structure (B).
The structure then converges more slowly in the energy 
as the atoms move toward the configuration in C. %
At this step (step \#13), 
the structure resembles that of the diamond %
(mean absolute discrepancy of $\sim$ 0.17 Bohr per DOF),
and we obtain an AFQMC total energy that is about 1 part in 3,100 higher than the 
global minimum at the diamond structure. 
Convergence is considered reached at this step for the first stage,
and the optimization undergoes a few steps around this converged position,
with a position averaging performed among these converged steps \cite{SC_GeoOpt_paper}
to yield a new starting position for the next stage,
as indicated by the green oval and arrow.  
By refining the optimization in a second stage of SET, with smaller targeted statistical error in the AFQMC force computations and
a reduced step size in FSSD,  %
we approach the correct minimum diamond structure as depicted in 
D. %
The SOAP similarity kernel~\cite{Bartok_SOAP_2013} $(1-K^{\mathrm{SOAP}})$
is a measure of how similar the structure is to the target. %
Our final structure in D has
a SOAP similarity kernel difference %
of $10^{-12}$ (mean absolute discrepancy of $\sim$ 0.011 Bohr per DOF),
and a total energy within one statistical error bar %
or one part in 106,000 of the energy of the ideal diamond structure.

\begin{figure}
\includegraphics[width=1\linewidth]{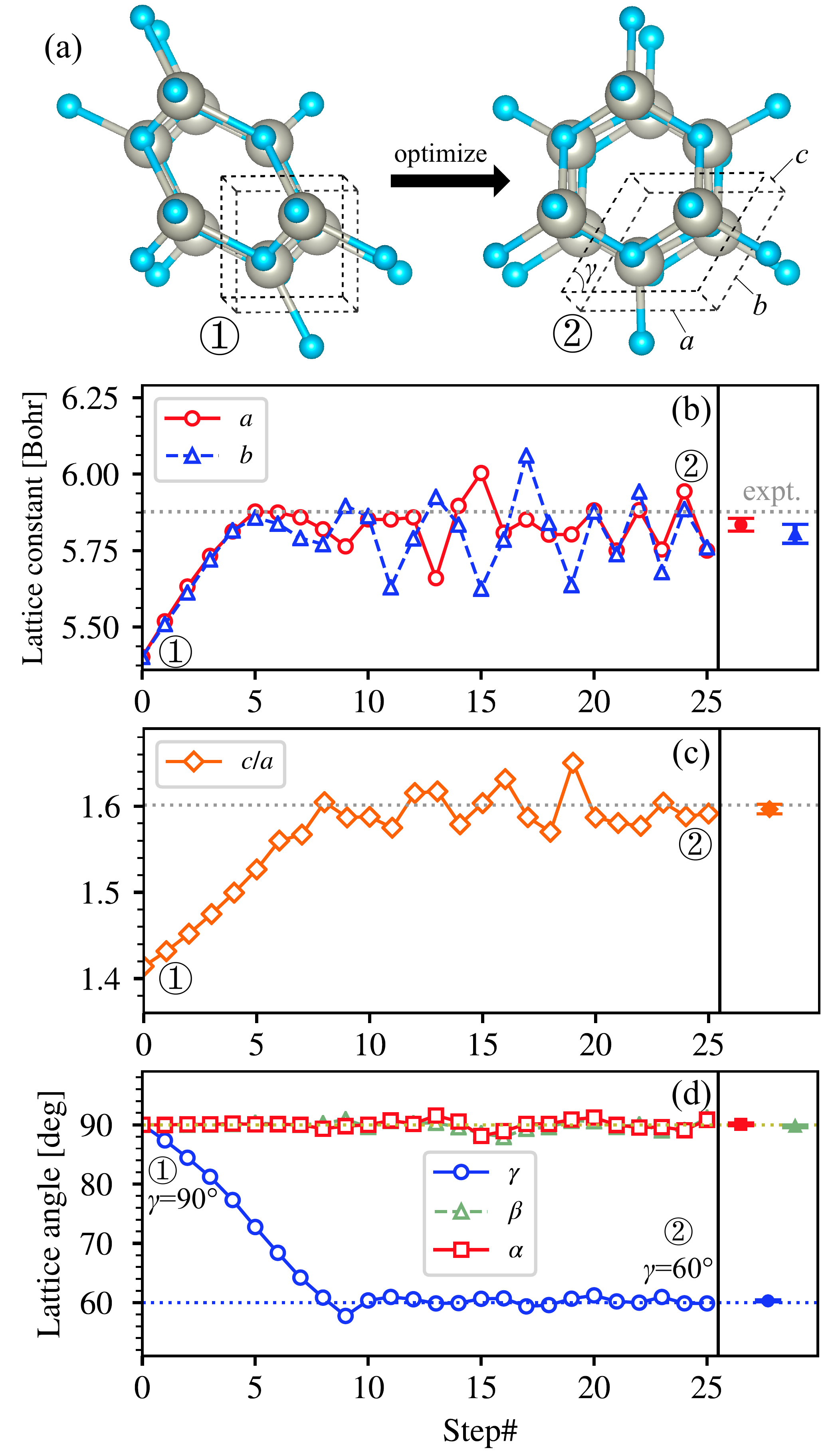}\caption{\label{fig:stress-FSSD}
Optimization of \CHANGED{the} lattice volume and shape in solid AlN.
(a) shows the initial and target structures.
(b)(c)(d) show the lattice constant
$a,b$, the ratio $c/a$, and the 3 lattice angles $\alpha,\beta,\gamma$,
respectively. Convergence is reached at step 10. On the right side
of each plot, the average from step 10 to step 25 is shown with the estimated statistical error bar.
Experimental values are shown in dotted lines for comparison.
}
\end{figure}

In the second example, we optimize %
the lattice volume and shape in solid AlN in the wurtzite structure.
Fig.~\ref{fig:stress-FSSD}(a) illustrates the setup.
The fractional atomic positions in the cell are  %
fixed to be the values of the wurtzite ($\mathrm{P6_{3}mc}$) structure. The initial
structure has a %
mismatch between the atom positions and the lattice structure,
which is %
tetragonal supercell of a cubic NaCl lattice
($c=7.64$ Bohr, $a=b=\sqrt{2}c/2$).
The target structure, which is the global minimum under ambient condition,
is the
wurtzite lattice shown on the right. 
This optimization procedure involves  6 degrees of freedom:
the lattice constants ($a,b$) and the lattice shape ($c/a$;
$\alpha,\beta,\gamma$).
We again apply the $\mathrm{FSSD}\times\mathrm{SET}$ algorithm for the optimization.
Instead of the forces as in the example above,
this requires repeated computations of the  
stress tensors with the PW-AFQMC algorithm outlined in the previous section. 
Fig.~\ref{fig:stress-FSSD}(b)(c)(d) demonstrate
how the lattice %
structure transforms towards the %
global minimum.
Convergence
of all DOF is seen %
at step \#10 with one
stage of $\mathrm{FSSD}\times\mathrm{SET}$. %
The evolution into a final structure of hexagonal wurtzite lattice is evident:
$c/a$ increases from $\sqrt{2}$ to $\sim1.60$, and $\gamma$ changes
from $90\lyxmathsym{\textdegree}$ to $60\lyxmathsym{\textdegree}$.
The averaged lattice parameters after convergence show very good agreement
with experimental results.

\section{\label{sec:conclusion}Conclusion and Outlook}

We have presented a method for accurate computations of %
interatomic forces and
stress tensors  %
in solid state systems, under the PW-AFQMC framework. %
The approach is outlined in detail, with a discussion of the sources of errors.
Benchmark calculations were performed using accurate total energies to test the formalism and implementation of the direct 
computation under the Hellmann-Feynman scheme. The approach is then applied in two simple solids as examples, demonstrating fully 
ab initio structural optimizations of both atomic positions and lattice structures. 

The work paves the way for structural optimizations in realistic materials with an accurate many-body method.  
This opens exciting new opportunities for more predictive computations in correlated materials.
A number of questions remain to be further explored to allow systematic applications, including reducing finite-size effects,
quantifying the accuracy in strongly correlated materials,  improving 
computational efficiency in our formalism, exploring the BP approach versus automatic differentiation, etc.

Interatomic forces are also key ingredients for computation of the phonon spectrum. The availability of forces from the approach we have presented thus makes possible many-body 
computation of phonon spectra in solids. 
A crucial new ingredient which enables systematic phonon calculations is the use of correlated sampling  \cite{SC_Phonon_paper},
which allows  estimates of small differences of systems in proximity, or derivatives by finite-difference. 
When combined with the approach presented in this work, 
we can then efficiently compute the derivatives of forces and stresses. %
A recent improvement of the correlated sampling algorithm has introduced population control \cite{SC_CorrSamp_paper},
which significantly improves its efficiency and effectiveness.

\begin{acknowledgments}
We are grateful to F.~Ma for help and for providing the pseudopotentials used in the present study. 
We thank H. Krakauer, M. Motta, F. Ma, M. A. Morales, L. K. Wagner, and S. Sorella
for useful discussions. 
S.C. thanks the Center for Computational Quantum
Physics, Flatiron Institute for support and hospitality. We also acknowledge
support from the U.S. Department of Energy (DOE) under Grant No. DE-SC0001303.
The authors thank William \& Mary Research Computing and Flatiron
Institute Scientific Computing Center for computational resources
and technical support. The Flatiron Institute is a division of the
Simons Foundation.
\end{acknowledgments}

\appendix
\numberwithin{equation}{section}

\section{The electron-ion force term}\label{app:psp-force}

We provide some additional details on the differentiation of the  pseudopotential coefficients. 
The local pseudopotential coefficient, $v_\mathrm{ei}^\mathrm{L}$, is given by:
\begin{equation}
v_\mathrm{ei}^\mathrm{L}(\mathbf{Q}) = \frac{1}{\Omega}\sum_i \mathcal{V}_{i}^\mathrm{L}(|\mathbf{Q}|) e^{-i \vec{\tau_i}\cdot{\mathbf{Q}}}\,,
\label{eq:psp-L-coeff}
\end{equation}
where $i$ loops over atoms, $\mathcal{V}^\mathrm{L}(Q)$ is a function
interpolated from the pseudopotential %
(we use multiple-projector norm-conserving pseudopotentials), 
and is the same for atoms of the same species. 
Differentiating this with respect to an atomic position $\vec{\tau}_{\mu}$ %
involves multiplying $i \mathbf{Q} \delta_{i\mu}$ to each term of the sum.
Since plane wave AFQMC uses convolutions instead of matrix multiplications, 
an additional Fourier transform to real space is performed and
saved for repeated use.

The nonlocal pseudopotential coefficient, $v_\mathrm{ei}^\mathrm{NL}$, is given by the Kleinman-Bylander form~\cite{KleinmanBylander}: %
\begin{equation}
v_\mathrm{ei}^\mathrm{NL}(\mathbf{G},\mathbf{G}') = \sum_J \frac{1}{\eta_J} u^\star_{J,\mathbf{G}} u_{J,\mathbf{G}'}\,,
\label{eq:psp-NL-coeff}
\end{equation} %
where $J$ loops over ``projectors'' and represents a combination of $\{i,l,m\}$, $i$ is the atom number and $l,m$ are the azimuthal and magnetic quantum numbers, $\eta_J$ is a constant for each $J$, and
\begin{equation}
u_{J,\mathbf{G}} = \frac{4\pi}{\sqrt{\Omega}}e^{i\vec{\tau}_i\cdot\mathbf{G}^\mathbf{k}} \mathcal{V}_{J}^\mathrm{NL}(|\mathbf{G}^\mathbf{k}|)Y^{*}_{l,m}(\mathbf{G}^\mathbf{k}) \,,
\label{eq:psp-NL-U}
\end{equation} %
where $\mathbf{G}^\mathbf{k}$ is a short hand for $\mathbf{G}+\mathbf{k}$
($\mathbf{k}$ is the twist angle for a twisted boundary condition).
$Y^{*}_{l,m}$ are \textit{complex-conjugated} spherical harmonics taking the polar coordinates angle $(\theta,\varphi)$ of the input vector.

Differentiating $v_\mathrm{ei}^\mathrm{NL}$ %
creates two terms. In each of them,
one of the $u_{J,\mathbf{G}}$ is unchanged,
while the other will be multiplied by $-i\mathbf{G}^\mathbf{k} \delta_{i,\mu}$:
\begin{equation}
\begin{aligned}
- \frac{\partial v_\mathrm{ei}^\mathrm{NL}(\mathbf{G},\mathbf{G}')}{\partial \tau_{\mu a}} =  \sum_J \frac{i\delta_{i\mu}}{\eta_J}  [& (u^\star_{J,\mathbf{G}}G_{a}) u_{J,\mathbf{G}'}\\
 - & u^\star_{J,\mathbf{G}} (G'_{a} u_{J,\mathbf{G}'})]\,,
\end{aligned}
\end{equation}
where $\tau_{\mu a}$ denotes the coordinate in the $a$-direction of the $\mu$-th atom.

Unlike the local electron-ion force, its nonlocal counterpart is not computed with convolutions. 
However, by writing the pseudopotential in the Kleinman-Bylander form,
the dimension has already been drastically reduced. 
Using the notation $U$ to represent the matrix of $u_{J,\mathbf{G}}$, and $\mathfrak{U}_a$ to represent the matrix of $(G_a u_{J,\mathbf{G}})$,
we group $U^\dagger$ or $\mathfrak{U}_a^\dagger$ with $\Phi^\dagger$, and $U$ or $\mathfrak{U}_a$ with $\Theta$,
and compute the matrix multiplication within each group first.
Sums on $J$ and all electrons are then performed, where $\delta_{i\mu}$ takes effect. In summary, one computes

\begin{equation}
\sum_{t\in\mathrm{electrons}} \sum_J \frac{i\delta_{i\mu}}{\eta_J}
[
( \mathfrak{U}_a \Phi )_{tJ}^\dagger ( U \Theta )_{Jt} - ( U \Phi )_{tJ}^\dagger ( \mathfrak{U}_a \Theta )_{Jt}
]\,,
\end{equation}
\noindent
where $U$ and $\mathfrak{U}_a$ are matrices of dimensions $(J,\mathbf{G})$, $\Theta$ and $\Phi$ are matrices of dimensions $(\mathbf{G},t)$.

\section{The electron-ion stress term}\label{app:psp-stress}

Based on the formulae in Appendix~\ref{app:psp-force}, we can also compute the electron-ion contribution to the stress, for which we now have to consider the dependency on $\mathbf{G},\mathbf{Q},\Omega$ as well.
For the local part:
\begin{equation}
-\frac{1}{\Omega}\frac{\partial v_\mathrm{ei}^\mathrm{L}(\mathbf{Q})}{\partial\epsilon_{ab}}=
\frac{1}{\Omega^2}\sum_i [\dot{\mathcal{V}}^\mathrm{L}_i(|\mathbf{Q}|)\frac{Q_a Q_b}{|\mathbf{Q}|}+\delta_{ab} \mathcal{V}^\mathrm{L}_j(|\mathbf{Q}|)]e^{-i\vec{\tau}_i\cdot\mathbf{Q}} \,,
\end{equation}
where %
$\dot{\mathcal{V}}^\mathrm{L}(Q) \equiv\mathrm{d}\mathcal{V}^\mathrm{L}(Q)/\mathrm{d}Q$ is obtained
by taking direct derivative of the cubic spline function used for interpolation. 
This entire object can be pre-computed and used to replace $v_\mathrm{ei}^\mathrm{L}$
in the energy computation routine to obtain the local pseudopotential stress contribution.
For the nonlocal part,
\begin{equation}
\begin{aligned}
-\frac{1}{\Omega}\frac{\partial v_\mathrm{ei}^\mathrm{NL}(\mathbf{G},\mathbf{G}')}{\partial\epsilon_{ab}}=
\sum_{J,\mathbf{G},\mathbf{G}'}\frac{1}{\eta_J}[&
(\bar{u}_{J,\mathbf{G};ab})^\star u_{J,\mathbf{G}'} \\
&+(u_{J,\mathbf{G}})^\star\bar{u}_{J,\mathbf{G}';ab} ]\,,
\end{aligned}
\end{equation}
where $\bar{u}_{J,\mathbf{G};ab}$ is a shorthand for $(-1/\Omega)(\partial u_{J,\mathbf{G}}/\partial\epsilon_{ab})$, and contains three terms:

\begin{enumerate}

\item A contribution from $\Omega^{-1/2}$, which is just $(\delta_{ab}/2\Omega) \times u_{J,\mathbf{G}}$.

\item A contribution from the derivative of $\mathcal{V}^\mathrm{NL}_J(G)$:
\begin{equation}
\frac{4\pi}{\Omega^{3/2}}e^{i\mathbf{x}_i\cdot\mathbf{G}^\mathbf{k}} \frac{G^\mathbf{k}_a G^\mathbf{k}_b}{|\mathbf{G}^\mathbf{k}|} [\mathcal{V}_{J}'(|\mathbf{G}^\mathbf{k}|)\cdot Y^{*}_{l,m} (\mathbf{G}^\mathbf{k}) ]\,.
\end{equation}

\item A contribution from the derivative of the spherical harmonics,
\begin{equation}
\frac{4\pi}{\Omega^{3/2}}e^{i\mathbf{x}_i\cdot\mathbf{G}^\mathbf{k}} [\mathcal{V}_{J}(|\mathbf{G}^\mathbf{k}|)\cdot \frac{\partial Y^{*}_{l,m}}{\partial G^\mathbf{k}_a}G^\mathbf{k}_b ]\,,
\end{equation}
which is computed %
together with the  spherical harmonics themselves,
and 
can be obtained %
with any library that computes $(\partial Y_{l,m}/\partial \theta)$
and $(\partial Y_{l,m}/\partial \varphi)$,
with a coordinate transformation from $(G,\theta,\varphi)$ to $(G_x,G_y,G_z)$.

\end{enumerate}

Written in full, for the nonlocal electron-ion stress, one computes:
\begin{equation}
\sum_{t\in\mathrm{electrons}} \sum_J \frac{1}{\eta_J}
[
( \bar{U}_{ab} \Phi )_{tJ}^\dagger ( U \Theta )_{Jt} + ( U \Phi )_{tJ}^\dagger ( \bar{U}_{ab} \Theta )_{Jt}
]\,,
\end{equation}
where $\bar{U}_{ab}$ represents the matrix of $\bar{u}_{J,\mathbf{G};ab}$.

\bibliography{ForceStress_Paper_v1.3}

\end{document}